\documentclass[aps,superscriptaddress,amsmath,amssymb,floatfix,twocolumn,showpacs,amsfonts,longbibliography,reprint,showkeys]{revtex4-2}

\usepackage{times}
\usepackage[varg]{txfonts}
\usepackage{textcomp}
\usepackage{graphicx}
\usepackage{subfigure}
\usepackage{color}
\usepackage[colorlinks=true,citecolor=blue,urlcolor=blue,linkcolor=blue,hyperindex]{hyperref}
\usepackage{braket}
\usepackage{float}
\usepackage{xcolor}
\usepackage{overpic}
\usepackage{bm}
\usepackage{amsbsy}
\usepackage{array}
\usepackage{booktabs}
\usepackage{tabularx}

\begin{document}

\title{Superconductivity of bilayer two-orbital Hubbard model for La$_{3}$Ni$_{2}$O$_{7}$ under high pressure}

\author{Wei-Yang Chen}
\thanks{These two authors contributed equally to this work.}
\affiliation{State Key Laboratory of Optoelectronic Materials and Technologies, Guangdong Provincial Key Laboratory of Magnetoelectric Physics and Devices, Institute of Neutron Science and Technology, School of Physics, Sun Yat-Sen University, Guangzhou 510275, China}
\affiliation{School of Physical Sciences, Great Bay University, Donguan, 523000, China}
\author{Cui-Qun Chen}
\thanks{These two authors contributed equally to this work.}
\affiliation{State Key Laboratory of Optoelectronic Materials and Technologies, Guangdong Provincial Key Laboratory of Magnetoelectric Physics and Devices, Institute of Neutron Science and Technology, School of Physics, Sun Yat-Sen University, Guangzhou 510275, China}
\author{Meng Wang}
\affiliation{State Key Laboratory of Optoelectronic Materials and Technologies, Guangdong Provincial Key Laboratory of Magnetoelectric Physics and Devices, Institute of Neutron Science and Technology, School of Physics, Sun Yat-Sen University, Guangzhou 510275, China}
\author{Shou-Shu Gong}
\email[Corresponding author:]{shoushu.gong@gbu.edu.cn}
\affiliation{School of Physical Sciences, Great Bay University, Donguan, 523000, China}
\affiliation{Great Bay Institute for Advanced Study, Donguan, 523000, China}

\author{Dao-Xin Yao}
\email[Corresponding author:]{yaodaox@mail.sysu.edu.cn}
\affiliation{State Key Laboratory of Optoelectronic Materials and Technologies, Guangdong Provincial Key Laboratory of Magnetoelectric Physics and Devices, Institute of Neutron Science and Technology, School of Physics, Sun Yat-Sen University, Guangzhou 510275, China}

\begin{abstract}
By combining density functional theory (DFT) and density matrix renormalization group calculations, we investigate the unusual pressure dependence of superconducting transition temperature ($T_c$) in the nickelate superconductor La$_{3}$Ni$_{2}$O$_{7}$.
Using the parameters obtained by fitting the DFT results, we map a quantum phase diagram of a bilayer two-orbital Hubbard model with increasing pressure.
Near $3/8$-filling, we find a strong spin density wave (SDW) order, which at $3/8$-filling shows a real-space pattern similar to the spin-charge stripe order. 
At $21/64$-filling, we find a superconducting phase with interlayer superconductivity (SC) in both the $d_{z^2}$ and $d_{x^2-y^2}$ orbitals, as well as in-plane SC in the $d_{z^2}$ orbital.
Intriguingly, the SC is weakened with increasing pressure and transitions to a Luttinger liquid above $80$ GPa, which qualitatively agrees with the experimental observations of decreasing $T_c$ with increasing pressure and a transition to Fermi liquid above $80$ GPa.
We further show that the ratio of interaction to hopping integral, which reduces moderately with increasing pressure, may play a dominant role in the weakening of SC.
Our results not only find a robust SC through suppressing the competing SDW order, but also give new insight into the unusual pressure dependence of SC in La$_{3}$Ni$_{2}$O$_{7}$.
\end{abstract}

\keywords{Superconductivity, Nickelate superconductor, Spin density wave, DMRG, High-pressure}

\maketitle

\section{Introduction}\label{sec:introduction}
The pursuit of unconventional superconductivity (SC) has long stood at the forefront of condensed matter physics. 
The recently found nickel oxide superconductors~\cite{Li2019, PhysRevLett.125.027001,Sun2023,Zhu2024,Ko2025,Zhou2025} have sparked a surge of experimental and theoretical studies. 
The multilayered structures in nickel oxides open a new route for the exploration of unconventional pairing mechanism.

The bulk compound La$_{3}$Ni$_{2}$O$_{7}$, which has a bilayer two-orbital structure, exhibits a SC with a high transition temperature $T_c \approx 80$ K under a pressure of $14$ GPa~\cite{Sun2023}.
The average valence state of nickel (Ni) atoms in La$_{3}$Ni$_{2}$O$_{7}$ is Ni$^{2.5+}$ ($3d^{7.5}$), and the $d_{z^2}$ and $d_{x^{2}-y^{2}}$ orbitals are nearly half-filled and quarter-filled, respectively~\cite{Sun2023,PhysRevLett.131.126001}. 
Although theoretical studies have found that the interlayer coupling between the $d_{z^2}$ orbitals is crucial for driving SC in La$_{3}$Ni$_{2}$O$_{7}$, the pairing symmetry and mechanism remain highly controversial~\cite{PhysRevLett.131.126001, PhysRevB.108.L140505, Luo_2024, PhysRevLett.131.236002,PhysRevLett.132.036502, PhysRevB.109.045154,PhysRevB.110.024514, Jiang_2024,cpl_40_12_127401, Xia2025,PhysRevB.111.L180508,Schlomer2024,171w-6kjw,yang2025evolutionintralayerinterlayersuperconductivity}.
Due to the complex interplay among orbital hybridization, Hund's coupling, and correlation, both the interlayer $s_\pm$-wave pairing symmetry~\cite{PhysRevLett.131.126001, PhysRevB.108.L140505, Luo_2024, PhysRevLett.131.236002,PhysRevLett.132.036502, PhysRevB.109.045154} and in-plane $d$-wave pairing symmetry~\cite{PhysRevB.110.024514, Jiang_2024,cpl_40_12_127401, Xia2025,PhysRevB.111.L180508} have been proposed based on different considerations of the dominant pairing orbital and driving force.

Very recently, a high-pressure experiment up to $104$ GPa for La$_{3}$Ni$_{2}$O$_{7}$ revealed an unusual pressure dependence of $T_c$~\cite{10.1093/nsr/nwaf220}, which may give new clues for clarifying the nature of its SC.
In this experiment, $T_c$ monotonically decreases with increasing pressure above $18$ GPa, and the SC gives way to a Fermi liquid
above $80$ GPa~\cite{10.1093/nsr/nwaf220}.
This finding has stimulated broad theoretical interests.
Based on a bilayer two-orbital Hubbard model, combined density functional theory (DFT) and functional renormalization group (FRG) calculations find that increasing pressure can weaken spin fluctuations that mediate the $s_{\pm}$-wave pairing, and thus $T_c$ decreases with pressure~\cite{PhysRevLett.134.076001}.
Another strong-coupling perspective based on a bilayer two-orbital $t$-$J$ model suggests that increased pressure can strengthen orbital hybridization, which induces orbital competition and therefore suppresses SC~\cite{171w-6kjw}. 
Currently, the Hubbard repulsion in La$_{3}$Ni$_{2}$O$_{7}$ is considered to be of intermediate strength, which may be more challenging for FRG calculation and go beyond the scope of $t$-$J$ model.
Thus, the origin of this weakening of SC by pressure remains an outstanding challenge in the study of La$_{3}$Ni$_{2}$O$_{7}$, which calls for further accurate study in the experimentally relevant parameter regime.

\begin{figure*}[htbp]
\centering
\includegraphics[width=1\textwidth]{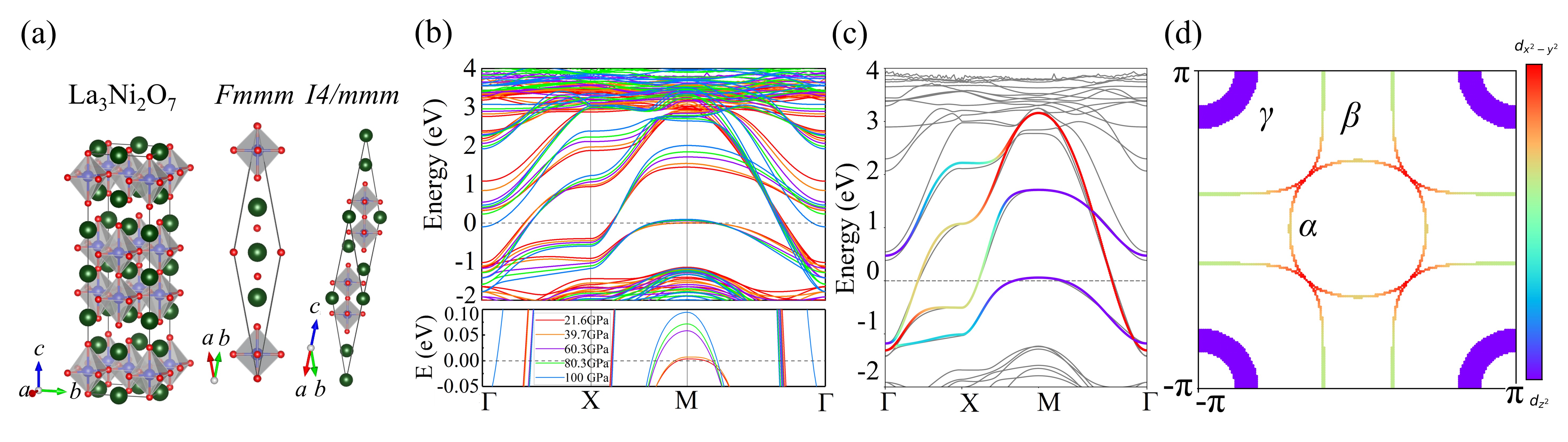}
\caption{(a) Crystal structures of La$_3$Ni$_2$O$_7$. The middle and right panels denote the primitive cells of $Fmmm$ and $I4/mmm$ phases. (b) DFT calculated band structures of primitive cell La$_3$Ni$_2$O$_7$ under various pressures. The bottom panel shows a zoom in of the flat bonding band ($\gamma$) near the Fermi level at the $M$ corner of the Brillouin zone. (c) Comparison of the DFT band structure (gray) and the fitted band structure from TB model (colored) under the pressure of $60.3$ GPa. (d) Fermi surface with one hole pocket ($\gamma$) and two electron pockets ($\alpha$ and $\beta$) of the TB model at $60.3$ GPa. The color bar denotes the orbital weight of the $d_{x^2-y^2}$ and $d_{z^2}$ orbitals.}
\label{fig1}
\end{figure*}

\begin{table*}[t]
\caption{Hopping parameters and on-site energies of the bilayer two-orbital TB model under different pressures. Here, $t^{\mu\nu}_{[lmn]}$ denotes the hopping integral connected by the $[0,0,0]$-$[l,m,n]$ bond between orbitals $\mu$ and $\nu$. $\mu, \nu = x$ and $z$ denote the $d_{x^2-y^2}$ and $d_{z^2}$ orbitals, respectively.
The unit of hopping integrals and on-site energies are eV. }
\label{tab1}
\noindent\begin{centering}
\begin{tabular}{ccccccccccccccc}
\hline \hline 
 Pressure (GPa) & Space group & $\epsilon_x$ & $\epsilon_z$ & $t^{xx}_{[100]}$ & $t^{zz}_{[100]}$ & $t^{xx}_{[110]}$ & $t^{zz}_{[110]}$ & $t^{xx}_{[200]}$ & $t^{zz}_{[200]}$ &$t^{xz}_{[100]}$ & $t^{xz}_{[200]}$ & $t^{zz}_{[001]}$ & $t^{xz}_{[101]}$ \tabularnewline
\hline 
21.6 & $Fmmm$ & 0.891 & 0.358 & -0.493 & -0.127 & 0.066 & -0.027 & -0.071 & -0.022 & 0.240 & 0.039 & -0.678 & -0.021 \tabularnewline
39.7 & $Fmmm$ & 0.900 & 0.345 & -0.522 & -0.140 & 0.065 & -0.023 & -0.073 & -0.018 & 0.253 & 0.040 & -0.724 & -0.026\tabularnewline
60.3 & $I4/mmm$ & 0.975 & 0.428 & -0.559 & -0.155 & 0.066 & -0.028 & -0.079 & -0.012 & 0.282 & 0.042 & -0.826 & -0.038 \tabularnewline
80.3 & $I4/mmm$ & 0.995 & 0.465 & -0.597 & -0.169 & 0.064 & -0.032 & -0.072 & -0.016 & 0.310 & 0.038 & -0.876 & -0.030 \tabularnewline
100.0 & $I4/mmm$ & 1.072 & 0.534 & -0.642 & -0.162 & 0.063 & -0.020 & -0.095 & -0.005 & 0.294 & 0.049 & -0.933 & -0.055 \tabularnewline
\hline \hline 
\end{tabular}
\par\end{centering}
\end{table*}

In addition, density wave orders in La$_{3}$Ni$_{2}$O$_{7}$ are also elusive~\cite{Seo1996, PhysRevLett.131.206501, PhysRevLett.131.206501, Liu2022, Liu2024,Liu2022, chen2024e,ZHAO20251239,XIE20243221,Meng2024}, which may constitute an important piece of the puzzle in the nature of SC.
At ambient pressure, La$_{3}$Ni$_{2}$O$_{7}$ undergoes a density wave transition characterized by resistance kinks around $110$ K and $153$ K~\cite{PhysRevB.63.245120, PhysRevX.14.011040, Khasanov2025}.
Below $\sim 150$ K, $\mu$SR, NMR and RIXS measurements have reported signatures of a spin density wave (SDW) order with the wavevector ($\pi$/2, $\pi$/2, $\pi$)~\cite{ZHAO20251239, Chen2024, Khasanov2025, chen2024e, ZHAO20251239}, with two candidates proposed: the spin-charge stripe order with alternating lines of magnetic moments and non-magnetic stripes, and the double spin stripe order with alternating double-parallel magnetic moment stripes~\cite{Chen2024}. 
However, the precise magnetic structure remains under debate.
With applied pressure, the SDW transition temperature increases, but the resistance kink temperature decreases, suggesting two density wave transitions and another possible charge density wave (CDW) order~\cite{Khasanov2025}.
To establish a comprehensive understanding of the electronic properties of La$_{3}$Ni$_{2}$O$_{7}$, it is also highly desired to clarify the nature of the density wave orders and their interplay with SC.

In this work, we explore the pressure dependence of SC as well as the interplay between density wave order and SC in La$_{3}$Ni$_{2}$O$_{7}$ based on a bilayer two-orbital Hubbard model. 
We first adopt DFT to study the electronic structures of La$_{3}$Ni$_{2}$O$_{7}$ using the lattice constants measured by experiment~\cite{10.1093/nsr/nwaf220}, giving the parameters of the tight-binding (TB) model under different pressures.
The correlation effects are further considered by the unbiased density matrix renormalization group (DMRG) calculations on a ladder geometry.
Taking the hopping integrals and on-site potentials of the TB model, and considering an intermediate Hubbard repulsion and a Hund's coupling, we map a quantum phase diagram of the system with increasing pressure $21.6 - 100$ GPa.
We further examine the electronic filling per unit cell (corresponding to the two Ni atoms in the bilayer) from $\eta_e = 3/8$ to $5/16$, where the electron filling is defined as the ratio of electron number to the total electron number of fully occupied orbitals.
This variation corresponds to doping approximately $0.25$ holes per Ni ion, which may be achieved experimentally via La substitution by Pr, Sr, or similar rare-earth elements~\cite{Zhou2025, Hao2025}.

For $\eta_e = 3/8 - 11/32$, we find a density wave phase with a strong SDW order in both $d_{z^2}$ and $d_{x^2-y^2}$ orbitals, but without a static CDW.
In particular, at $3/8$ filling the SDW exhibits a spin configuration similar to the spin-charge stripe order along a lattice direction. 
At lower $\eta_e = 21/64$, we find a superconducting phase with interlayer SC in both orbitals and in-plane SC in the $d_{z^2}$ orbital, accompanied by suppression of the SDW.
Interestingly, with increasing pressure, the SC is gradually weakened and the system shows a transition to a Luttinger liquid (LL) at about $80$ GPa, which qualitatively agrees with the experimental observation of decreasing $T_c$ in La$_{3}$Ni$_{2}$O$_{7}$.

By examining the variation of model parameters with increasing pressure, we find small changes in electron density and the ratios between hopping integrals but a moderate decrease in the ratio of interaction to hopping, which is shown to make an important contribution for the weakening of SC by a comparative study.
Our results provide an understanding for the experimental observation of the bulk La$_{3}$Ni$_{2}$O$_{7}$ based on a bilayer two-orbital Hubbard model and suggest a path to SC through weakening the competing SDW order.

\section{ Band structure and tight-binding model}
Our DFT calculations are performed by Vienna ab initio simulation package (VASP)~\citep{VASP1,VASP2}, in which the projector augmented wave (PAW) \citep{PAW1,PAW2} method with a 600 eV plane-wave cutoff is applied. 
The generalized gradient approximation (GGA) of Perdew-Burke-Ernzerhof form (PBE) exchange correlation potential is adopted \citep{PhysRevLett.77.3865}. 
The convergence criterion of force is set to be 0.001\ eV/{\rm \AA}and total energy convergence criterion is set to be $10^{-7}$ eV. 
The crystal structures of La$_3$Ni$_2$O$_7$ measured under high pressure are shown in Fig.~\ref{fig1}(a). 
Up to $104$ GPa, La$_3$Ni$_2$O$_7$ shows two structural transitions: from $Amam$ to $Fmmm$ space group at $12.3$ GPa, and subsequently from $Fmmm$ to $I4/mmm$ at $46.8$ GPa~\citep{10.1093/nsr/nwaf220}. 
A $\Gamma$-centered $19\times19\times19$ Monkhorst Pack k-mesh grid is used for primitive cell of $I4/mmm$ phase and a $\Gamma$-centered $14\times14\times14$ Monkhorst Pack k-mesh grid for primitive cell of $Fmmm$ phase. 
In structural relaxations, we adopt the experimental refined lattice constants of $\mathrm{La_{3}Ni_{2}O_{7}}$ ~\citep{10.1093/nsr/nwaf220}, and optimize the atomic coordinates of the system. 
In the $Fmmm$ structure, the in-plane lattice parameters $a$ and $b$ are unequal.
With increasing pressure, the difference between $a$ and $b$ gradually decreases.
At $46.8$ GPa, a tetragonal structure with $a = b$ stabilizes.
To account for the correlation effect of Ni atoms, DFT+U method is employed and $U_{eff}$ is set to 3.5 eV \cite{Yang2024,PhysRevB.57.1505}. 
To obtain the projected tight-binding models, we further perform Wannier downfolding as implemented in WANNIER90~\citep{w90} package, in which the good convergence is achieved.

The obtained band structures are presented in Fig.~\ref{fig1}(b), where the bands around the Fermi level are contributed from the Ni-$e_g$ states. 
As pressure increases, the bandwidth of the Ni-$e_g$ sector broadens, accompanied by an enlargement of the energy gap between the bonding ($\gamma$) and antibonding ($\gamma_{an}$) bands.
The bottom panel of Fig.~\ref{fig1}(b) is the zoom in near the Fermi level, showing that the $M$ corner of the $\gamma$ band in the $Fmmm$ structure is flatter than that in the $I4/mmm$ structure, which thus gives rise to a larger density of states around the Fermi level. 
With increasing pressure from $21.6$ to $100$ GPa, the $\gamma$ band shifts upward and results in the larger $\gamma$ pocket in the $M$ corner.

Based on the Wannier downfolding on the Ni-$e_g$ orbitals, we construct a bilayer two-orbital TB model: 
\begin{eqnarray}
\label{eq1}
H_0=\sum_{i,\mu,\sigma}\epsilon_{\mu}\hat{c}^{\dagger}_{i\mu\sigma}\hat{c}_{i\mu\sigma}+\sum_{i,j,\mu,\nu,\sigma}t_{ij}^{\mu\nu}(\hat{c}^{\dagger}_{i\mu\sigma}\hat{c}_{j\nu\sigma}+h.c),
\end{eqnarray}
where $i/j$, $\sigma$ and $\mu/\nu$ denote the indexes of site (for both layers), spin, and orbital, respectively.
$\epsilon_{\mu}$ represents the on-site energy of the orbital $\mu$ ($\mu,\nu = x$ and $z$ denote the $d_{x^2-y^2}$ and $d_{z^2}$ orbitals, respectively).
These parameters obtained by fitting the DFT band structures are presented in Table~\ref{tab1}, and the TB band structures agree well with the DFT results, as shown in Fig.~\ref{fig1}(c) and Supplementary Information (SI). 
The obtained bilayer two-orbital model gives three pockets ($\alpha$, $\beta$ and $\gamma$) on the Fermi surface, consistent with previous studies~\cite{PhysRevLett.131.126001}. 
The flat hole pocket $\gamma$ around the $M$ corner is mainly characterized by the $d_{z^2}$ orbital, and the electron pockets $\alpha$ and $\beta$ are dominated by the mix of $d_{z^2}$ and $d_{x^2-y^2}$ orbitals (see Fig.~\ref{fig1}(d)). 
Along the diagonal direction ($M-\Gamma$ path), the inter-orbital hoppings disappear due to the requirement of crystal symmetry. 
As a consequence, the bands along the diagonal direction can be described solely by the $d_{z^2}$ or $d_{x^2-y^2}$ orbital. 
The $d_{z^2}$ orbitals comprise the bonding and antibonding states, which are separated by an energy splitting $2|t_{[001]}^{zz}|$ at the $M$ point (see the definition of $t_{[001]}^{zz}$ in the caption of Table~\ref{tab1}).
As pressure increases, the nearest-neighbor (NN) hopping integrals are monotonically enhanced, especially by $30\%$ for $t_{[100]}^{xz}$ and $37.6\%$ for $t_{[001]}^{zz}$, characterizing the strengthened in-plane orbital hybridization and interlayer hopping between the $d_{z^2}$ orbitals, respectively.
With the structural transition from $Fmmm$ to $I4/mmm$, the on-site energies $\epsilon_{\mu}$ also show a drastic change.

\section{Phase diagram obtained by DMRG calculation}

\begin{figure}[htbp]
\centering
\includegraphics[width=0.5\textwidth]{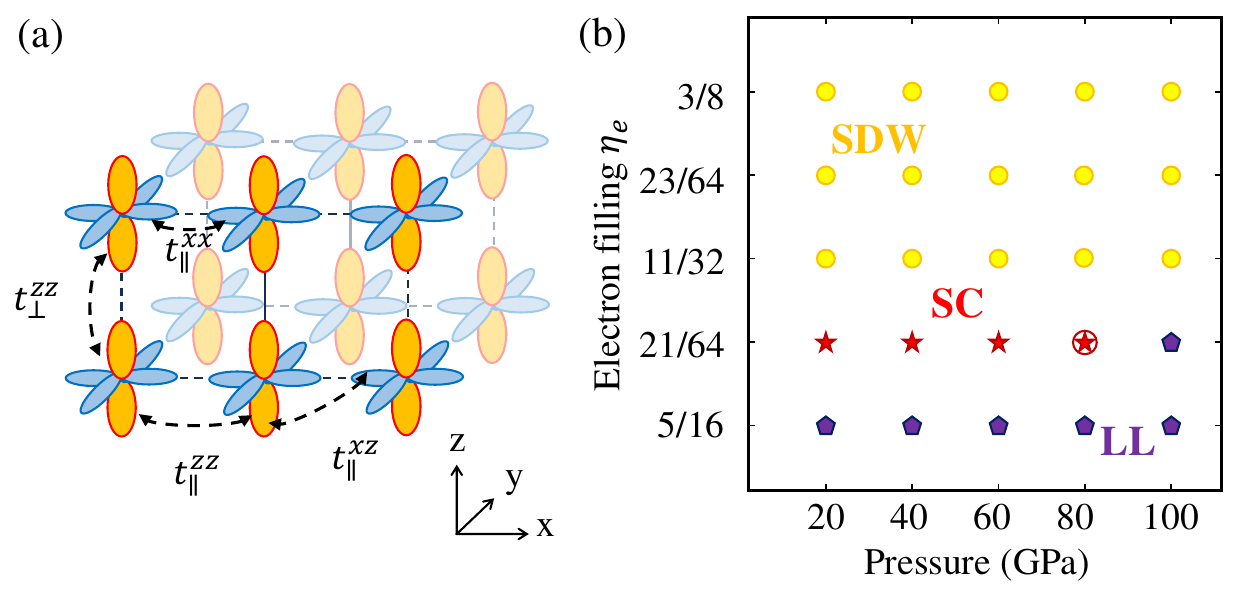}
\caption{(a) Schematic figure of the bilayer two-orbital model with the $d_{z^2}$ and $d_{x^2-y^2}$ orbitals. The hopping integrals $t_{||}^{zz}$, $t_{||}^{xz}$, $t_{||}^{xx}$, and $t_{\perp}^{zz}$ are chosen from the TB model shown in Table~\ref{tab1}. (b) DMRG phase diagram of the model on a ladder geometry with system width $L_y=1$ and the interactions $U = 4.0$ eV and $J_H = 0.5$ eV. By tuning electron filling per unit cell $\eta_e$ and pressure, the system exhibits a spin density wave (SDW) phase (yellow circle), a superconducting phase (red star), and a Luttinger-liquid (LL) phase (purple pentagon). At $\eta_e = 21/64$, the system has a transition from SC to LL near $80$ GPa (red star in a circle). The LL phase may evolve to a Fermi-liquid phase in the two-dimensional limit.}
\label{fig2}
\end{figure}

\begin{figure*}[htbp]
\centering
\includegraphics[width=1\textwidth]{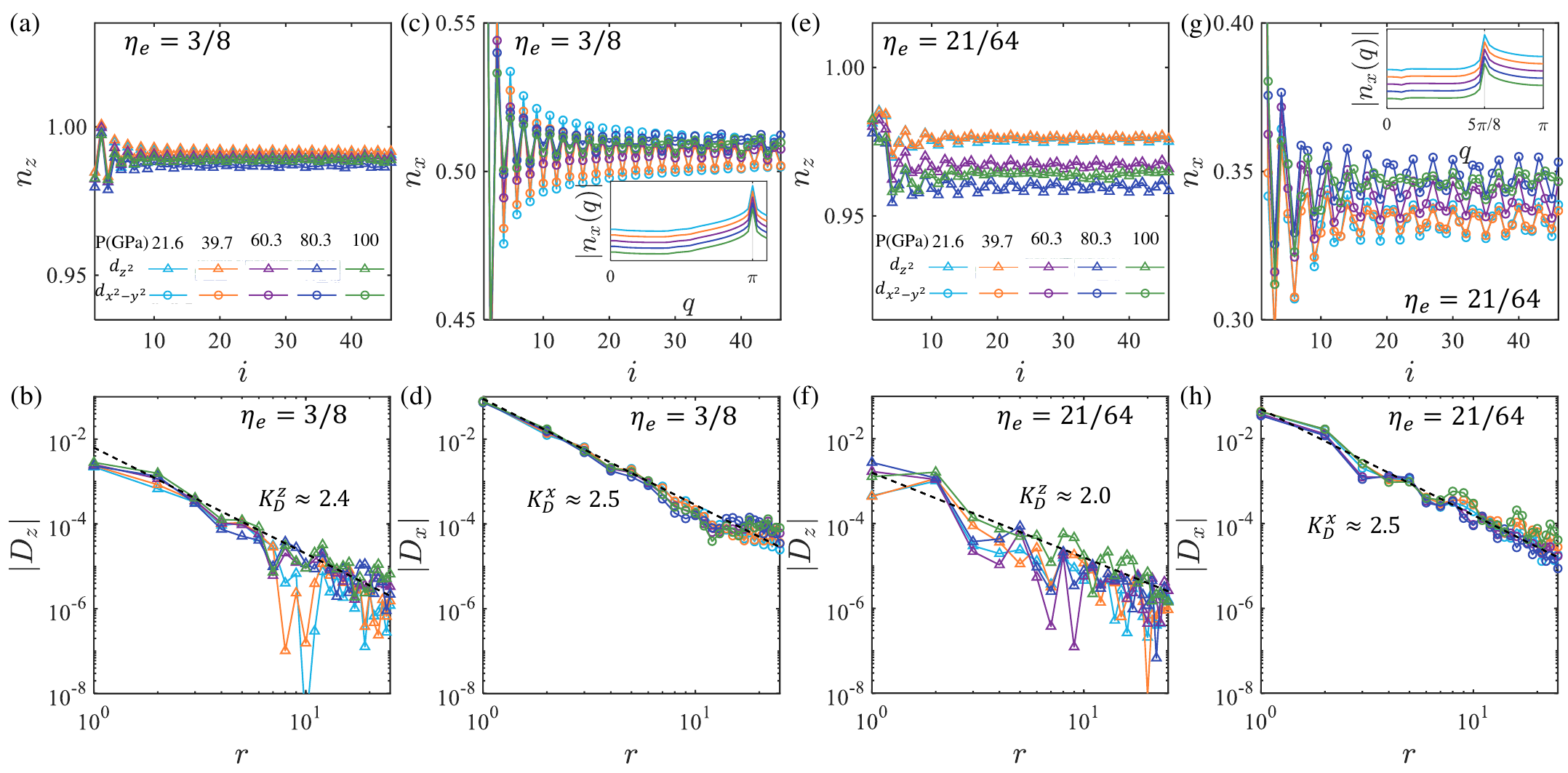}
\caption{Electron density distributions $n_\mu$ (a, c, e, g ) and density correlation functions $D_\mu$ (b, d, f, h) of both $d_{z^{2}}$ (triangle) and $d_{x^{2}-y^{2}}$ (circle) orbitals for the electron filling $\eta_e = 3/8$ and $21/64$ under different pressures. The insets in (c) and (g) show the Fourier transform of electron density distribution in the $d_{x^{2}-y^{2}}$ orbital. $K^{\mu}_D$ ($\mu = z$ or $x$) denotes the obtained power exponents by the algebraic fitting of density correlation functions.}
\label{fig3}
\end{figure*}

\noindent Based on the TB model Eq.~\eqref{eq1}, we further consider the correlation effects using the DMRG calculation.
Since the NN hopping integrals are dominant in Table~\ref{tab1}, we only consider the NN hoppings and study a bilayer two-orbital Hubbard model $\hat{H} = \hat{H}_{kin} + \hat{H}_{int}$ as depicted in Fig.~\ref{fig2}(a)~\cite{PhysRevLett.131.126001}. 
We set the NN in-plane hopping between the $d_{x^2-y^2}$ ($d_{z^2}$) orbitals as $t_{\parallel}^{xx} = t^{xx}_{[100]}$ ($t_{\parallel}^{zz} = t^{zz}_{[100]}$), and the interlayer hopping between the $d_{z^2}$ orbitals as $t_{\perp}^{zz} = t^{zz}_{[001]}$.
For the in-plane orbital hybridization, the sign arises from the difference in the phase sign of the $d_{x^2-y^2}$ orbital wave functions across their spatial distributions. In our DMRG simulation of the system with lattice width $L_y=1$, we choose the chain direction along the crystal $a$-axis, and accordingly set a positive value for the hybridization, i.e., $t_{\parallel}^{xz} = t_{\parallel}^{zx} = t^{xz}_{[100]}$.
For simplicity, we redefine the on-site energies $\varepsilon_{z} = 0$ and $\varepsilon_{x} = \epsilon_{x} - \epsilon_{z}$.
Thus, the kinetic energy can be written as
\begin{align}
\hat{H}_{\text{kin}} &= \sum_{\langle ij \rangle, l, \mu, \nu, \sigma} t_{\parallel}^{\mu \nu} 
\left( \hat{c}_{i,l,\mu,\sigma}^{\dagger} \hat{c}_{j,l,\nu,\sigma} + \text{h.c.} \right) \notag \\
&\quad + \sum_{i,\sigma} t_{\perp}^{zz} 
\left( \hat{c}_{i,1,z,\sigma}^{\dagger} \hat{c}_{i,2,z,\sigma} + \text{h.c.} \right) \notag \\
&\quad + \sum_{i,l,\mu,\sigma} \varepsilon_{\mu} \hat{n}_{i,l,\mu,\sigma},
\label{eq2}
\end{align} 
where the subscript $l$ is the layer index and $i,j$ mark the site in the given layer. 
The interaction part is given as
\begin{align}
\hat{H}_{\text{int}} &= \sum_{i,l,\mu} U\, \hat{n}_{i,l,\mu,\uparrow} \hat{n}_{i,l,\mu,\downarrow} \notag \\
&\quad + \sum_{i,l,\sigma,\sigma'} \left( U' - J_H \delta_{\sigma,\sigma'} \right) 
\hat{n}_{i,l,x,\sigma} \hat{n}_{i,l,z,\sigma'}, 
\label{eq3}
\end{align}
where $U$, $U^{\prime}$ and $J_H$ are intra-orbital repulsion, inter-orbital repulsion, and Hund's coupling, respectively, with the relation $U^{\prime} = U - 2J_H$~\cite{PhysRevB.18.4945}. 
Since the interactions originate either from the electron cloud overlap of local orbital or that between distinct orbitals at the same site, these quantities should be less sensitive to pressure and have been considered as pressure independent in previous studies~\cite{PhysRevLett.134.076001}.
Here, we choose an intermediate $U = 4.0$ eV and $J_H = 0.5$ eV~\cite{PhysRevLett.131.126001, PhysRevLett.131.206501}.
Since the hopping integrals vary with pressure, we choose $U$ as the energy unit.
For example, at $21.6$ GPa, $t^{xx}_{\parallel}/U = -0.123$, $t^{zz}_{\parallel}/U = -0.032$, $t^{xz}_{\parallel}/U = 0.060$, $t^{zz}_{\perp}/U = -0.170$, and $\varepsilon_{x}/U = 0.133$.
Although SC has been found in the $t$-$J$ type models~\cite{PhysRevLett.132.036502, Schlomer2024,171w-6kjw,yang2025evolutionintralayerinterlayersuperconductivity} and some simplified Hubbard models~\cite{cpl_40_12_127401,PhysRevB.109.045154, PhysRevB.111.L180508}, obtaining a robust SC in a bilayer two-orbital Hubbard model in the experimentally relevant parameter regime is still an urgent task. 
In recent DMRG study of the bilayer two-orbital Hubbard model~\cite{cpl_40_12_127401,PhysRevB.109.045154, PhysRevB.111.L180508}, the Hund’s coupling has rarely been considered. Nevertheless, Hund’s coupling may potentially play a vital role in multi-orbital systems. 
Therefore, in this work we consider the Kanamori density-density terms encoding the Ising-type Hund’s coupling.
Notice that the interaction part in this model omits the spin-flip and onsite pair-hopping terms, which represent the fluctuating parts of the Hund's coupling. 
At the moderate Hund's coupling strength ($J_H/U=0.125$), these couplings are much weaker compared with the considered density-density interactions. 
Besides, given the computational costs of including these interactions for mapping the phase diagram, we leave the consideration of spin-flip and pair-hopping terms to future study.

We employ DMRG calculation~\cite{PhysRevLett.69.2863}, implemented using the iTensor library \cite{10.21468/SciPostPhysCodeb.4}, to determine the ground state of the system. 
Due to the computation limit, we consider the lattice width $L_y = 1$ and length up to $L_x = 64$, with the open boundary conditions in the $x$ direction.
Despite potential finite-size effects due to the limited size in the $y$-direction, our findings provide significant insights into the interplay between doping and pressure, as well as the competition between SDW and SC.
We implement the charge $U(1)$ and spin $U(1)$ symmetries. 
We use the bond dimensions up to $12000$, giving accurate results with small truncation errors in the order of $10^{-7}$.

\begin{figure*}[htbp]
\centering
\includegraphics[width=1\textwidth]{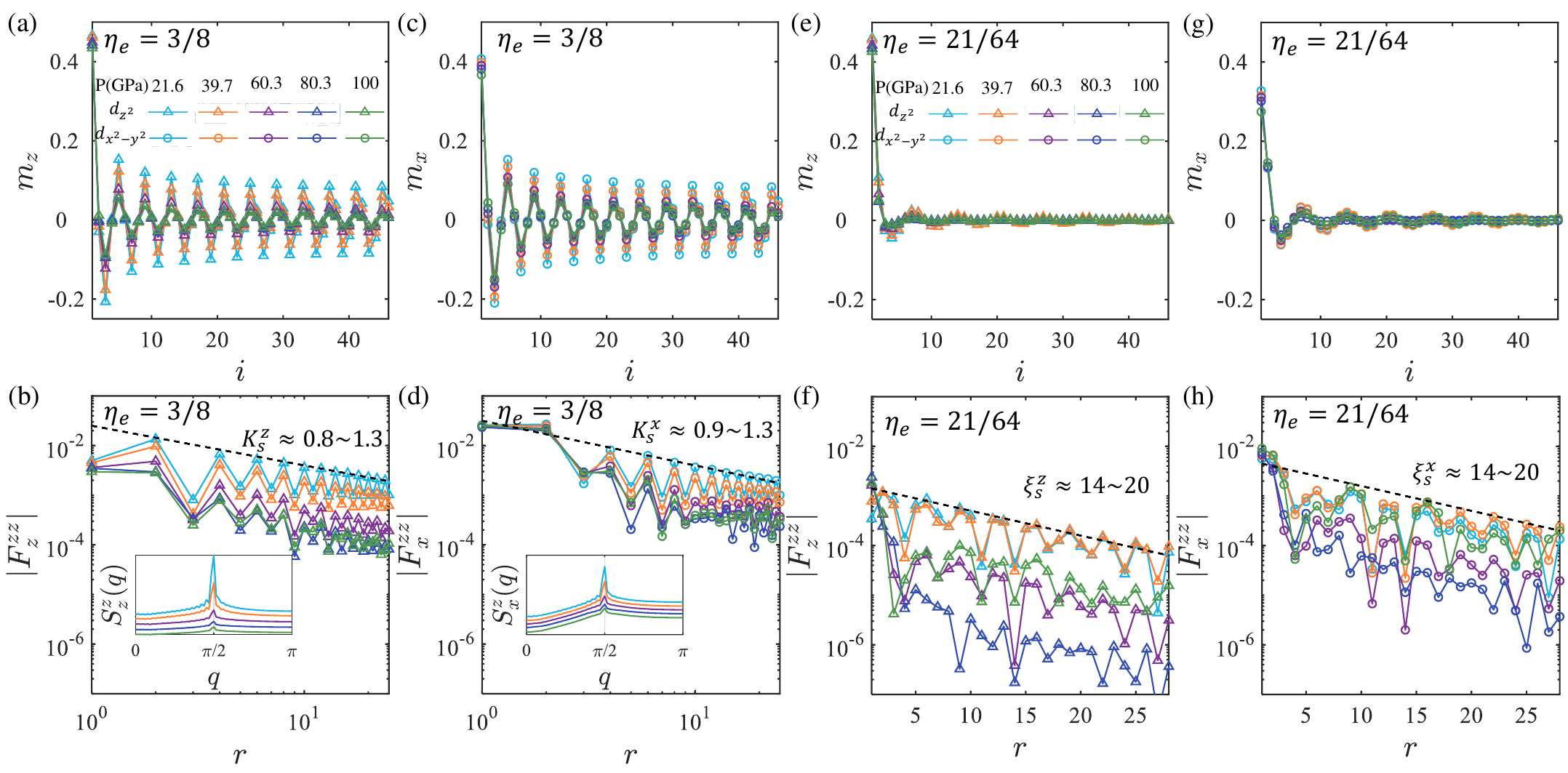}
\caption{Local magnetic moments $m_\mu$ (a, c, e, g) and spin correlation functions $F^{zz}_\mu$ (b, d, f, h) of both $d_{z^{2}}$ (triangle) and $d_{x^{2}-y^{2}}$ (circle) orbitals for the electron filling $\eta_e = 3/8$ and $21/64$ under different pressures. Edge pinning magnetic field is introduced to compute local magnetic moments. Spin correlation functions for $\eta_e = 3/8$ are plotted as double-logarithmic scale with the fitted power exponents $K^{\mu}_s$. For $\eta_e = 21/64$, spin correlations are presented as semi-logarithmic scale with the spin correlation lengths $\xi^{\mu}_s$. The insets in (b) and (d) show the corresponding spin structure factor at $\eta_e = 3/8$.}
\label{fig4}
\end{figure*}

By tuning pressure and electron filling, we obtain a quantum phase diagram as shown in Fig.~\ref{fig2}(b).
For $\eta_e = 3/8 - 11/32$, we find a nonsuperconducting phase with a strong SDW order in both orbitals but without a static CDW.
The spin and charge density waves in the $d_{x^2-y^2}$ orbital are intertwined, as evidenced by their connected wavevectors $2Q^{x}_S = Q^{x}_D$.
In particular, at $\eta_e = 3/8$ the SDW shows a wavevector $Q_S = \pi/2$ and appears to be similar to the spin-charge stripe order along a lattice direction.
With increasing doping, SDW is suppressed at $\eta_e = 21/64$ and a superconducting state emerges at the pressure of $21.6$ GPa, showing the interlayer SC in both orbitals and in-plane SC in the $d_{z^2}$ orbital as well.
This SC is characterized by the formed hole pairing and the algebraic pairing correlation functions with the power exponents $K^{\mu}_{SC} < 2$.
Intriguingly, the SC is gradually weakened by increasing pressure, which is characterized by increased $K^{\mu}_{SC}$, leading to the disappearance of SC above $80$ GPa, which qualitatively agrees with the experimental observation of decreasing $T_c$ in La$_{3}$Ni$_{2}$O$_{7}$~\cite{Sun2023, 10.1093/nsr/nwaf220}. 
At lower filling, the electrons in both orbitals behave like a LL with algebraic correlation functions.
\\

\subsection{Weak charge density wave}

To investigate the charge order, we compute the local charge density $n_{\mu}(i) = (1/2) {\textstyle \sum_{l,\sigma}} \left \langle \hat{n}_{i,l,\mu,\sigma}\right\rangle$
and density correlation function $D_{\mu}(r) = (1/2) \sum_{l} (\langle \hat{n}_{i,l,\mu} \hat{n}_{j,l,\mu} \rangle - \langle \hat{n}_{i,l,\mu} \rangle \langle \hat{n}_{j,l,\mu} \rangle)$, where $\hat{n}_{i,l,\mu} = \sum_{\sigma} \hat{n}_{i,l,\mu,\sigma} $ and $r = | i-j |$. 
In the SDW phase, the charge density profiles of both orbitals decay from the open boundary to the bulk [Fig.~\ref{fig3}(a) for $n_z(i)$ and Fig.~\ref{fig3}(c) for $n_x(i)$], characterizing the absence of a static CDW order.
While $n_z(i)$ is nearly uniform, $n_x(i)$ shows a power-law decay from the open boundary to the bulk, which behaves like a Friedel oscillation~\cite{white2002}.
As shown by the peak of the Fourier transform $n_\mu(q) = \sum_{j} e^{i q r_j} ( n_{\mu}(j) - \bar{n}_\mu )$ in the inset of Fig.~\ref{fig3}(c), where $\bar{n}_\mu$ is the averaged electron density of orbital $\mu$, $n_x(q)$ has a peak at $q = \pi$ for $\eta_e = 3/8$.
Meanwhile, both density correlation functions appear to follow an algebraic decay $D_{\mu}(r) \sim r^{-K^{\mu}_{D}}$ with $K^{\mu}_{D} \gtrsim 2$, identifying a weak quasi-long-range CDW.
The charge order in both orbitals remains weak in the SC [Figs.~\ref{fig3}(e)-\ref{fig3}(h)] and LL regimes in SI, with the power exponents $K^{\mu}_{D} \gtrsim 2$.
In the inset of Fig.~\ref{fig3}(g), we notice that the peak of $n_x(q)$ shifts to $Q^x_D = 5\pi/8$ at $\eta_e = 21/64$.
By examining the results at other fillings, we find that $Q^x_D$ is always proportional to the average electron density, i.e., $Q^x_D \propto \bar{n}_x \pi$ in SI.

In addition, we also notice the featured dependence of the electron densities on the total filling and pressure.
With increasing hole doping, $\bar{n}_x$ decreases significantly but $\bar{n}_z$ only decreases slightly, indicating that the holes are mainly doped in the $d_{x^{2}-y^{2}}$ orbital, which is consistent with the larger on-site energy as shown in Table~\ref{tab1}.
With increasing pressure for each filling, a small number of electrons are transferred from $d_{z^{2}}$ to $d_{x^{2}-y^{2}}$ orbital.
Similar charge transfer by pressure has also been observed in the DFT results by the upward shift of the $\gamma$ band [Fig.~\ref{fig1}(b)] and the decrease of the density of states in the $d_{z^{2}}$ orbital~\cite{Huo2025}.

\subsection{Spin density wave}
To study magnetic order, we compute the local magnetic moment and spin correlation function.
To reduce degeneracy and obtain stable magnetic moments, we introduce the edge pinning magnetic field coupled with spin-$z$ component~\cite{PhysRevLett.99.127004}.
Since these quantities behave consistently in the two layers, we present the results in the upper layer, i.e. $m_{\mu}(i) \equiv \langle \hat{S}^{z}_{i,1,\mu} \rangle$ and $F^{zz}_{\mu}(r) \equiv\langle \hat{S}^z_{i,1,\mu} \hat{S}^z_{i+r,1,\mu} \rangle$.

In the SDW phase, the magnetic moments in both orbitals show very slow power-law decay from the boundary to the bulk [Figs.~\ref{fig4}(a) and \ref{fig4}(c)], and the spin correlation functions decay algebraically with small power exponents $K^{\mu}_s \approx 1$ [Figs.~\ref{fig4}(b) and \ref{fig4}(d)], characterizing a strong quasi-long-range magnetic order in both orbitals.
The Fourier transformations of magnetic moment and spin correlation show a consistent SDW wavevector $Q^{\mu}_S$ (see the spin structure factor in the insets of Fig.~\ref{fig4}), which are the same for the two orbitals ($Q^{z}_S = Q^x_S$) possibly because of the Hund's coupling. 
In the $d_{x^2-y^2}$ orbital, the CDW and SDW wavevectors exhibit a fixed relation $Q^x_D = 2Q^x_S$, suggesting the intertwined charge and spin even though SDW is much stronger than CDW.
Intriguingly, for $\eta_e = 3/8$,$Q^{\mu}_S = \pi/2$ and the magnetic moments appear to better fit the spin-charge stripe order along a lattice direction.
Nevertheless, the magnetic structure is sensitive to electron filling since $2Q^x_S = Q^x_D \propto \bar{n}_x \pi$.
In this SDW phase, we also notice that the SDW is weakened by the increased pressure and hole doping.
In a recent DMRG study of a similar model at $3/8$ filling and the $J_H \rightarrow \infty$ limit, where the system can be reduced to an effective $t$-$J$ type model, a SDW with momentum $(\pi/2, 0)$ was found~\cite{yang2025magnetismsuperconductivitybilayernickelate}. 
This SDW order was suggested as a consequence of the competition between the $(\pi,\pi)$ N\'eel order driven by the super-exchange of the $d_{x^2-y^2}$ orbital and the double-exchange ferromagnetism driven by the kinetic energy~\cite{yang2025magnetismsuperconductivitybilayernickelate}. 
For future study, it would be interesting to tune the Hund's coupling to explore the connection between the SDW found in Ref.~\cite{yang2025magnetismsuperconductivitybilayernickelate} and this work, which would be helpful for addressing the origin of this SDW order.

\begin{figure*}[htbp]
\centering
\includegraphics[width=1\textwidth]{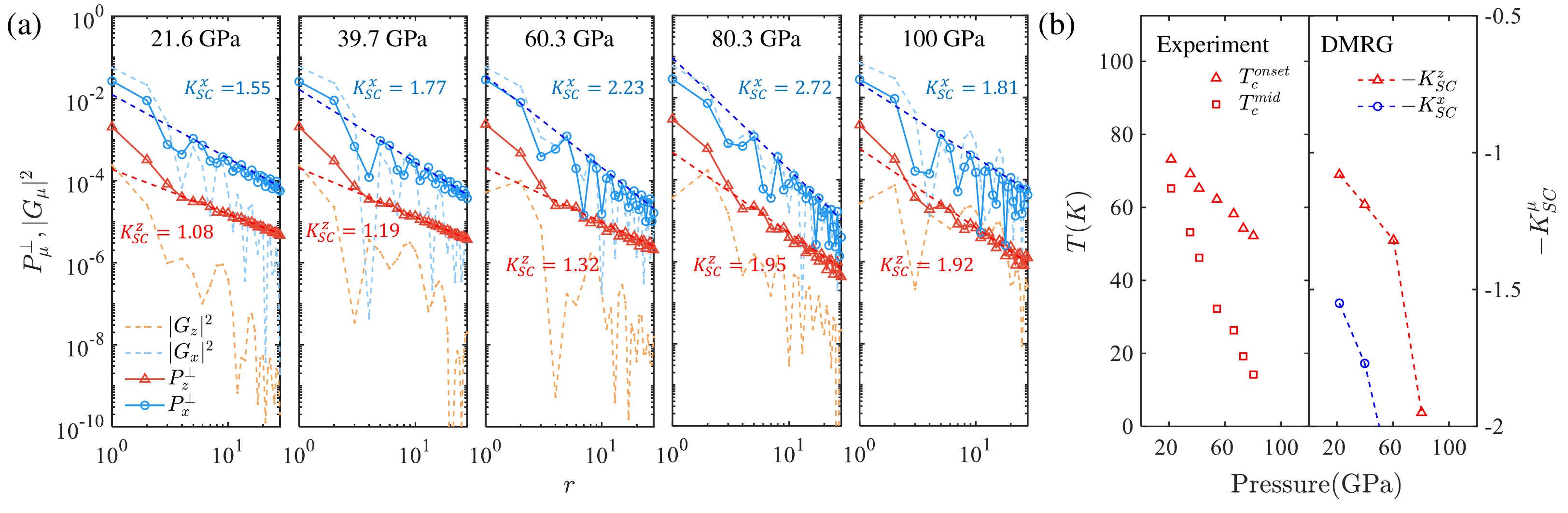}
\caption{(a) Comparisons of the interlayer pairing correlation function (solid) and the square of single-particle Green's function (dotted) in both $d_{x^{2}-y^{2}}$ (circular) and $d_{z^{2}}$ (triangle) orbitals under different pressures for $\eta_e = 21/64$. The interlayer pairing correlation functions $P^{\perp}_x(r)$ and $P^{\perp}_{z}(r)$ are fitted algebraically, giving the power exponents $K^x_{SC}$ and $K^{z}_{SC}$, respectively. (b) Comparison of SC transition temperature $T_c$ by the high-pressure resistance measurements~\cite{10.1093/nsr/nwaf220} and the power exponents $-K^{\mu}_{SC}$ obtained in subfigure (a).}
\label{fig5}
\end{figure*}

In the superconducting phase, both magnetic moments and spin correlations are significantly weakened [Figs.~\ref{fig4}(e)-\ref{fig4}(h)]. 
This suppression of magnetic order is consistent with the emergent SC because of the singlet pairing formed in both orbitals (see Fig.~\ref{fig5}).
In the LL phase, the spin correlation functions exhibit the power-law decay with $K^{\mu}_S \approx 2$ in SI. 
\\

\subsection{Superconductivity under high pressure}
To characterize SC, we measure both the interlayer and intralayer pairing correlations $P^{\perp}_\mu(r) = \langle \hat{\Delta }^{\perp\dagger}_{i,\mu} \hat{\Delta }^{\perp}_{i+r,\mu} \rangle$ and $P^{\parallel}_\mu(r) = (1/2) \sum_{l} \langle \hat{\Delta }^{\parallel \dagger}_{i,l,\mu} \hat{\Delta }^{\parallel}_{i+r,l,\mu} \rangle$, as well as the single-particle correlation functions $G_{\mu}(r) = (1/2) {\textstyle \sum_{l,\sigma}} \langle \hat{c}^{\dagger}_{i,l,\mu,\sigma} \hat{c}_{i+r,l,\mu,\sigma} \rangle$, where the interlayer and intralayer singlet annihilation operators of the orbital $\mu$ are defined as $ \hat{\Delta}^{\perp}_{i,\mu}=(\hat{c}_{i,1,\mu,\uparrow}\hat{c}_{i,2,\mu,\downarrow}-\hat{c}_{i,1,\mu,\downarrow}\hat{c}_{i,2,\mu,\uparrow} )/\sqrt{2}$ and $\hat{\Delta}^{\parallel}_{i,l,\mu} = (\hat{c}_{i,l,\mu,\uparrow}\hat{c}_{i+1,l,\mu,\downarrow} -\hat{c}_{i,l,\mu,\downarrow}\hat{c}_{i+1,l,\mu,\uparrow} )/\sqrt{2}$, respectively.
In previous studies, various pairing theories have been proposed, including interlayer $d_{z^2}$ orbital pairing with enhanced phase coherence from hybridization with metallic $d_{x^{2}-y^{2}}$ orbital~\cite{PhysRevB.108.L201108,PhysRevB.111.014512}, interlayer $d_{x^{2}-y^{2}}$ orbital pairing assisted by Hund's coupling and formed spin singlet between $d_{z^2}$ orbitals~\cite{PhysRevLett.132.146002}, as well as in-plane $d_{x^{2}-y^{2}}$ orbital pairing~\cite{PhysRevB.110.024514, Xia2025, Jiang_2024}.
In Fig.~\ref{fig5}(a), we show the interlayer pairing correlations and the squared single-particle correlations for both two orbitals at $\eta_e = 21/64$. 
At $21.6$ GPa, both $P^{\perp}_z(r)$ and $P^{\perp}_x(r)$ show a good power-law decay with small power exponents $K^z_{SC} \approx 1$ and $K^x_{SC} \approx 1.5$. 
Meanwhile, $P^{\perp}_z(r)$ and $P^{\perp}_x(r)$ are much stronger than $|G_z(r)|^2$ and $|G_x(r)|^2$, respectively.
These results unambiguously characterize the formed singlet pairing and a strong quasi-long-range interlayer SC.
For the in-plane SC, we find that it is absent in the $d_{x^{2}-y^{2}}$ orbital, but for the $d_{z^2}$ orbital $P^{\parallel}_{z}(r)$ also shows a relatively strong quasi-long-range order with the power exponent $\sim 1.12$ at $21.6$ GPa, seem SI.
Intriguingly, our DMRG results of this bilayer two-orbital Hubbard model find interlayer SC in both two orbitals and in-plane SC in the $d_{z^2}$ orbital as well.
Our DMRG results of the interlayer SC could agree with the predicted observations of the
two different pairing mechanisms: $d_{z^2}$-orbital pairing driven by hybridization~\cite{PhysRevB.108.L201108,PhysRevB.111.014512}, and $d_{x^2-y^2}$-orbital pairing mediated by Hund’s coupling~\cite{PhysRevLett.132.146002}. 
To further identify the dominant pairing mechanism in this model, we will separately tune the hybridization and Hund's coupling in future study. 

With increasing pressure, the interlayer SC in both orbitals are gradually weakened, characterized by the increase of power exponents $K^{\mu}_{SC}$. 
For the $d_{x^2-y^2}$ orbital, the interlayer SC appears to have already vanished at $60.3$ GPa, evidenced by the large power exponent $K^x_{SC} \approx 2.23$ and the very close magnitudes of $P^{\perp}_x(r)$ and $|G_x(r)|^2$.
Meanwhile, $K^{z}_{SC}$ continues to increase slightly from $21.6$ GPa to $60.3$ GPa but $|G_z(r)|$ maintains a fast exponential decay, indicating a weakened SC.
At $80.3$ GPa, $P^{\perp}_z(r)$ is strongly suppressed with $K^{z}_{SC} \approx 1.95$, which suggests that a transition may be occurring.
At $100$ GPa, $K^{z}_{SC} \approx 1.92$ and $|G_z(r)|^2$ is very close to $P^{\perp}_z(r)$, characterizing the vanished SC.
Concurrently, the in-plane SC in the $d_{z^2}$ orbital also disappears (see SI), showing a transition from SC to LL driven by increased pressure.
With growing system width, the LL phase may evolve to a Fermi-liquid phase in the 2D limit, which might be examined by other methods such as variational Monte Carlo or mean-field calculation.

Remarkably, the DMRG results of the pressure dependence of SC qualitatively agree with the experimental observation of decreasing $T_c$ in La$_3$Ni$_2$O$_7$.
In the left panel of Fig.~\ref{fig5}(b), we show the experimental data of $T^{onset}_c$ and $T^{mid}_c$ for La$_3$Ni$_2$O$_7$ with increasing pressure~\cite{10.1093/nsr/nwaf220}.
From $20$ GPa to $80$ GPa, the transition temperature of SC decreases monotonically, and the SC gives way to a Fermi liquid above $80$ GPa.
In the right panel of Fig.~\ref{fig5}(b), we plot the power exponents $-K^{\mu}_{SC}$ obtained in Fig.~\ref{fig5}(a), which also monotonically decrease with pressure and characterize the weakened SC.
Near $80$ GPa, a transition from SC to LL occurs.

To probe the origin of this pressure dependence of SC, we first consider the influence of charge density.
As shown in Figs.~\ref{fig3}(e) and \ref{fig3}(g), electrons are gradually transferred from $d_{z^2}$ to $d_{x^{2}-y^{2}}$ orbital with increasing pressure in the SC phase, but the change of charge density is nearly negligible.
For the $d_{z^2}$ orbital, $\bar{n}_z$ varies only from $0.976$ ($21.6$ GPa) to $0.960$ ($80.3$ GPa). 
As we have mentioned in Table~\ref{tab1}, some hopping integrals have a considerable growth with increasing pressure.
However, the relative ratios between these hoppings vary small, such as $t^{zz}_{\perp}/t^{xx}_{\parallel}$, $t^{xz}_{\parallel}/t^{xx}_{\parallel}$, and$t^{zz}_{\parallel}/t^{xx}_{\parallel}$, only around $6\% - 8.8\%$ from $21.6$ GPa to $80.3$ GPa.
On the other hand, when hopping integrals increase and interactions remain unchanged, the ratio of interaction to hopping decreases with increasing pressure.
For $\eta_e = 21/64$, $U/t^{xx}_{\parallel}$ decreases from $8.0$ at $21.6$ GPa to $6.6$ at $80.3$ GPa, which may play an important role in the weakening of SC.
To verify this conjecture, we further simulate a system at $21/64$ filling by choosing the hopping integrals of $21.6$ GPa as fixed and gradually turn down $U/t^{xx}_{\parallel}$ and $J_H/t^{xx}_{\parallel}$, which ignores the changes of the ratios between the hopping integrals.
Interestingly, we find a weakening of SC with decreasing $U/t^{xx}_{\parallel}$ and $J_H/t^{xx}_{\parallel}$, which is very similar to Figs.~\ref{fig5}(a)-\ref{fig5}(c) (see SI) and strongly suggests the ratio between hopping and interaction as the primary ingredient for the observations in Fig.~\ref{fig5}.

\section{Summary and discussion} 

We have investigated the unusual pressure dependence of SC transition temperature in La$_{3}$Ni$_{2}$O$_{7}$.
Using the hopping integrals and on-site energies by fitting the DFT band structures, we study a bilayer two-orbital Hubbard model with an intermediate Hubbard repulsion and a Hund's coupling by means of the unbiased DMRG simulation on a ladder geometry with system width $L_y = 1$. 
Our DMRG results seem to agree with the predictions of the two key pairing mechanisms proposed in recent theoretical studies~\cite{PhysRevB.108.L201108, PhysRevB.111.014512, PhysRevLett.132.146002}.
We map a quantum phase diagram by tuning the pressure and filling factor slightly below $3/8$.
Besides the SDW phase at larger filling and the LL phase at lower filling, at $\eta_e = 21/64$ we identify a superconducting phase with interlayer SC in both two orbitals and in-plane SC in the $d_{z^2}$ orbital as well.
The pairing correlations are weakened by increasing pressure, and the system shows a transition to LL near $80$ GPa, which remarkably agrees with the experimental observation of decreasing $T_c$~\cite{10.1093/nsr/nwaf220}.
By analyzing the change of model parameters, we find that the ratio of interaction to hopping integral reduces moderately with increasing pressure, and a further comparative study supports the important role of this factor in the weakening of SC.
In Ref.~\cite{PhysRevLett.134.076001}, the FRG calculations show that this reduced effective interaction with increasing pressure can weaken spin fluctuations and thus $T_c$ decreases.
Although our DMRG results do not explicitly characterize spin fluctuations, the observed essential role of the effective interaction agrees with the FRG conclusion. 
Notice that while the FRG calculations focus on $3/8$ filling~\cite{PhysRevLett.134.076001}, our results further unveil the competition between SDW and SC, as well as the importance of carrier density for the emergence of SC.

In the phase diagram Fig.~\ref{fig2}(b), the SDW is suppressed by increased hole doping and a SC emerges between the SDW and LL.
This picture is typical in doped correlated insulators such as cuprates~\cite{Ye2023}. 
With increasing pressure at $\eta_e = 21/64$, the increased ratio of hopping to interaction is also expected to suppress SC and give rise to a metallic state. 
In our comparative study, we have ignored the small variation of the ratios between hopping integrals, which might be more important near a metallic phase and thus may change the phase boundary between SC and LL. 
We will continue to study the impacts of orbital hybridization and Hund’s coupling on the competition between the different phases.

Note that in our phase diagram SC is absent at the experimentally relevant $3/8$ filling, but emerges at $21/64$ filling.
This result may originate from the finite-size effects or model simplification.
Although the $t$-$J$ type model with Hund's coupling can yield a SC at $3/8$ filling~\cite{171w-6kjw}, it may significantly underestimate the influence of charge fluctuations which should be important in the nickelates with an intermediate Hubbard repulsion~\cite{PhysRevLett.131.126001, PhysRevLett.131.206501,PhysRevLett.132.106002, Yang2024}.
For future study, considering additional interactions such as the spin-flip and pairing-hopping terms in this Hubbard type system, and increasing system size are highly desired to describe the material more accurately.
Interestingly, while the electron filling of the bulk La$_{3}$Ni$_{2}$O$_{7}$ is proposed to be $3/8$~\cite{Sun2023}, the optimal filling $21/64$ is closer to the filling proposed for the thin-film (La,Pr)$_{3}$Ni$_{2}$O$_{7}$~\cite{Zhou2025}.
More future studies with tuning electron filling and interaction may be crucial for clarifying the difference between the bulk and thin-film La$_{3}$Ni$_{2}$O$_{7}$.

\begin{acknowledgements}
The work at Sun Yat-sen University was supported by the National Natural Science Foundation of China ( 92565303, 12494591, 92165204), the National Key R$\&$D Program of China (2022YFA14028002), CAS Superconducting Research Project (SCZX-0101), Guangdong Provincial Key Laboratory of Magnetoelectric Physics and Devices (2022B1212010008), Research Center for Magnetoelectric Physics of Guangdong Province (2024B0303390001), and Guangdong Provincial Quantum Science Strategic Initiative (GDZX2401010). 
Wei-Yang Chen and Shou-Shu Gong were supported by the National Natural Science Foundation of China (12274014 and 12534009), the Guangdong Provincial Quantum Science Strategic Initiative (GDZX2501006), the Special Project in Key Areas for Universities in Guangdong Province (2023ZDZX3054), and the Dongguan Key Laboratory of Artificial Intelligence Design for Advanced Materials. 
The computational resources were partially supported by the SongShan Lake HPC Center (SSL-HPC) at Great Bay University.
\end{acknowledgements}

\section*{Conflict of interest}  
The authors declare that they have no conflict of interest.

\bibliography{references}

\end{document}